\documentclass[twocolumn,groupedaddress,
superscriptaddress,preprintnumbers,
amsmath,amssymb,
aps,
]{revtex4}
\usepackage{graphicx}
\usepackage{dcolumn}
\usepackage{bm}%
\usepackage[colorlinks=true,citecolor=blue]{hyperref}
\hypersetup{colorlinks=true,citecolor=blue,linkcolor=red,urlcolor=blue}

\usepackage{float}
\usepackage{amsmath}

\usepackage{color}
\usepackage{amsmath}
\usepackage{amsbsy}
\usepackage{color}
\usepackage{bbold}
\usepackage[colorlinks=true,citecolor=blue]{hyperref}
\DeclareGraphicsExtensions{.pdf,.eps,.png,.jpg,.mps}

\definecolor{textcolor}{cmyk}{0,0,0,1}
\definecolor{magenta}{rgb}{1,0,1}
\definecolor{green}{rgb}{0,1,0}
\definecolor{red}{rgb}{1,0,0}

\begin{document}
\newcommand{\be}{\begin{equation}}
\newcommand{\ee}{\end{equation}}
\newcommand{\bearr}{\begin{eqnarray}}
\newcommand{\eearr}{\end{eqnarray}}
\newcommand{\nn}{\nonumber}
\newcommand{\eps}{\varepsilon}

\title{Probing divacancy defects in a zigzag graphene nanoribbon through RKKY exchange interaction}
\author{Moslem Zare }
\email{mzare@yu.ac.ir}
\affiliation{Department of Physics, Yasouj University, Yasouj, Iran 75914-353, Iran}
\author{Reza Asgari}
	\affiliation{School of Physics, Institute for Research in Fundamental Sciences (IPM), Tehran 19395-5531, Iran}
	\affiliation{School of Nano Science, Institute for Research in Fundamental Sciences (IPM), Tehran 19395-5531, Iran}
	\affiliation{ARC Centre of Excellence in Future Low-Energy Electronics Technologies, UNSW Node, Sydney 2052, Australia}
\date{\today}

\begin{abstract}
 We investigate the effect of vacancy defects on the electronic and magnetic properties of zigzag graphene nanoribbons (zGNRs) by making use of the Green’s function formalism in combination with the tight-binding Hamiltonian.
The evolution of the indirect exchange coupling, known as Ruderman-Kittel-Kasuya-Yosida (RKKY) interaction, including single, double, and multiple 5-8-5 divacancy defects is explained.
Our numerical calculations show that the changes in the electronic structure and the exchange coupling of zGNRs depend significantly on the location of the divacancy defects with respect to the ribbon edges and on the number of the divacancy defects.
In the case both the impurities are located on the edge, the magnitude of the exchange coupling  is several orders of magnitude strengthen that result when they are placed on the interior of the nanoribbon.
Furthermore, a periodic divacancy causes a dramatic change in the magnetic ground state of the ribbon.
In the limit of high vacancy potential, the strength of the RKKY interaction is approximately independent of the Fermi energy.
\end{abstract}

\maketitle

\section{Introduction}\label{sec:Intro}
Graphene, a two-dimensional (2D) allotrope of carbon formed by a single layer of graphite~\cite{nov04}, has attracted great attention owing to the Dirac-like energy spectrum of its charge carriers and the resulting spectacular properties~\cite{gei07,cas09,voz10}.
Despite significant advancement in synthesis and processing of atomically precise graphene nanoribbons (GNRs), various structural defects generated during preparation are broadly considered to be inevitable.
The electronic, chemical, thermal, and mechanical properties of ideally 2D graphene are profoundly influenced by the presence of structural defects, lattice imperfections~\cite{gei07,cas09}, wrinkles, and even the ripples that always exist in graphene sheets~\cite{mey07,fas07}.
However, these ubiquitous extrinsic/intrinsic defects are considered the limiting factor for electronic transport in graphene through
charged impurities~\cite{Adam07,Tan07}, rippling~\cite{KatsnelsonPhilos08} or resonant scatterers~\cite{Titov,Z.H.Ni,Ferreira_prb2011} as well as give
rise to pseudo-magnetic gauge fields~\cite{voz10}.
Similar effects can be expected from topological lattice defects which spontaneously lead to corrugations in graphene~\cite{Seung}.
The ion-induced formation of lattice defects can be considered as a potential source of intervalley scattering in defected graphene that causes a diverging resistivity at low temperature, indicating the insulating behavior of graphene~\cite{JHChenPRL2009}.

As one of the intrinsic point defects in graphene, the vacancy has recently received enormous attention because of great implications on graphene devices~\cite{Kushmerick-Topsakal,hashimoto_nature2004,Skrypnyk,RobinsonPRL10,SkrypnykPRB10,wehlingPRB07,WehlingPRL10,Yuan2010,Wakabayashi_Jpn05,Cazalilla07,A.K.Mitchell13,Yazyev2007,S.Z.Liang12,Bilteanu}.  Such defects can be used to tailor or improve the physical characteristics of graphene~\cite{Kis04,Silva05,Nair,Yazyev2007,ugeda_prl2012,JHChenPRL2009,Chen_nature,Wang_prl2008,Stampfer2008} and generate unusual phenomena~\cite{Wakabayashi_Jpn05,Cazalilla07,A.K.Mitchell13,Yazyev2007,S.Z.Liang12,Bilteanu}.
For example, these vacancies cause the so-called resonant scattering at the Dirac points and are considered as a source of limiting graphene conductivity~\cite{Ferreira_prb2011}. In particular, vacancy defects are predicted to change the semimetallic nature of graphene to metallic behavior~\cite{Peres_prb2006}.

The vacancy defects can be artificially created in graphene by electron or ion irradiation~\cite{Nair,Yazyev2007} and visualized with atomic resolution by high-resolution transmission electron microscopy (HR-TEM)~\cite{hashimoto_nature2004} and scanning probe microscopy~\cite{Kelly96,Ruffieux2000}.
As a simple controllable and scalable method, acid treatment is known the simplest approach for vacancy defect creation~\cite{col08}.
The most probable form of defects generated by ion irradiation are single vacancies~\cite{ugeda_prl2012} that give rise to magnetic moments in single layer graphene~\cite{Nair,Yazyev2007}.

It has been shown that GNRs possess a finite bandgap~\cite{Fujita_Jpn1996,son_prl2006,Wakabayashi2010AdvMater} owing to the quantum confinement and edge effects, and promises to be suitable for development of realistic graphene-based nanodevices such as transistors~\cite{Tsengbook,Marmolejo}, thermoelectric generators~\cite{Sevincli,Dollfus,Tran17} and optoelectronic applications~\cite{Alavi,JZhu18,Soavi}.

Vacancies in a ribbon are the common type of defects that can influence the electronic characteristics of ribbons such as band gap and conductivity~\cite{Wang_prl2008,Stampfer2008}. For instance, the vacancy state of graphene could be strongly modified in the presence of edges in semi-infinite graphene sheets~\cite{Deng_prb2014}.
Therefore, they should be taken into account in graphene-based practical applications~\cite{Wang_prl2008,Stampfer2008}.
Many theoretical studies mostly focus on the effect of vacancies on the electronic properties of GNRs with their device characteristics~\cite{Rutter2007,Pereira_prl2006}.

Divacancies (DVs), type of defect created either by the coalescence of two single-vacancies or by removing two neighboring carbon atoms, may naturally appear as a stable defect during growth or can be created on purpose by electron or ion irradiation~\cite{hashimoto_nature2004,kotakoski_prl2011,kim_prb2011,robertson_nc2012,ugeda_prl2012,ugeda2}. It has been shown that the DVs in graphene is energetically favored over the monovacant defects because the formation energy of the DVs is lower than that of two isolated monovacancies and they have a tendency to coalesce to a DV~\cite{Lee2005prl,Krasheninnikov}. Moreover, the monovacancies can coalesce to form DV of its reconstruction without dangling bond~\cite{Borisova2013} and are the most important ones regarding the changes in transport properties~\cite{Ajayan,Navarro2005nat} a low concentration, about 0.03  orders of magnitude.

{\it Ab initio} calculations of dynamics and stability of DVs in graphene show that a DV is one of the most abundant defects in irradiated graphene~\cite{wang,Lehtinen,KrasheninnikovCPL} which have various reconstructed structures, such as triple pentagon-triple heptagon (555-777) and pentagon-octagon-pentagon (5-8-5) patterns~\cite{Kim2011prb,col08}.
A double-vacancy (5-8-5) defect is formed by the removal of two carbons, leaving a surface with two pentagonal rings and one octagonal ring~\cite{Lee2005prl}. On the other hand, a Stone–Wales defect is the rearrangement of the six-membered rings of graphene into two five (pentagons) and two seven (heptagons) rings. This rearrangement is a result of $\pi/2$ rotation of a C–C bond~\cite{hashimoto_nature2004}.

The formation of reconstructed DVs close to the edges of the Zigzag graphene nanoribbons (zGNRs) can be a practical way to make them partially ferromagnetic~\cite{Jaskolski}. This effect takes place even though the Dvs are produced by removing two atoms from opposite sublattices, which were balanced before reconstruction to 5-8-5 defects. There is a strong interaction between the defect-localized and edge bands which mix and split away from the Fermi level.
According to Lieb's theorem~\cite{lieb}, one way to attain  ferromagnetic graphene nanostructures is to impose sublattice imbalance. For instance, graphene systems with vacancies~\cite{palacios_jfr_prb2008,palacios_jfr_bry_fertig_sst2010,Pereira_prl2006,lps} have a non-zero spin due to the sublattice imbalance.
When these defects are present in zGNRs, they give rise to localized states and consequently can lead to spin effects and ribbon magnetization~\cite{palacios_jfr_prb2008,oeiras_prb2009}. Moreover, it has been found that zGNRs have spin-polarized edges, antiferromagnetically coupled in the ground state with total spin zero~\cite{Jaskolski2011prb} and can be ferromagnetic due to the presence of reconstructed 5-8-5 defects DVs near one edge.

In semiconducting armchair ribbons and two-dimensional graphene without global sublattice imbalance, there is a maximum defect density above which local magnetization disappears~\cite{palacios_jfr_prb2008}.
Overall, zGNRs with 5-8-5 DV defects may show either zero spin polarization~\cite{topsakal_prb2008} or spin-polarized transport in ribbons with narrow widths~\cite{oeiras_prb2009}.
For pristine graphene with a low concentration of vacancy defects, when the Fermi energy lies in the energy region where the density of states (DOS) is linear, the indirect exchange coupling, known as Ruderman-Kittel-Kasuya-Yosida (RKKY) interaction~\cite{Ruderman, Kasuya, Yosida} mediated by a background of conduction electrons of the host material, decay as $R^{-3}$~\cite{Habibi}. Otherwise, the presence of vacancies pushes the exponent ${-3}$ towards more negative values. A few percent vacancies wash out both atomic-scale oscillations and the Friedel oscillations due to the Fermi surface for a remarkable range of chemical potential values~\cite{Habibi}.

Having said earlier, the chemical, the mechanical, thermal and mechanical properties of GNRs, in the presence of 5-8-5 DVs, are studied extensively~\cite{hashimoto_nature2004,kotakoski_prl2011,kim_prb2011,robertson_nc2012,ugeda_prl2012,ugeda2,Barbary,Jaskolski}. However, the indirect exchange coupling for graphene nanoribbons in the presence of DVs has not been systematically reported.
Motivated by the effect of DVs on the electrical~\cite{Botello13,Zhao14PLA,Lherbier,Ren2010}, magnetic~\cite{Jaskolski,Petrovic} properties of GNRs, in this paper, we explain the evolution of the indirect exchange coupling in zGNRs. Within the tight-binding (TB) model we exploit the Green's function formalism to reveal the RKKY interaction between two magnetic impurities (MIs) placed on a zGNRs.

Our calculations show that the changes in the electronic structure and the exchange coupling of zGNRs depend on the location of the DVs with respect to the ribbon edges and on the number of the DVs. Introducing vacancies into zGNR changes the spatial variation of the RKKY interaction, particularly those magnetic moments located around the vacancies. We show that different values of the vacancy potential in the same zGRN give rise to different changes in the electronic and magnetic properties.
A periodic DV created in a zGNR causes a dramatic change in the magnetic ground state of the ribbon.
It is established that the presence of DVs in zGNR leads to the remarkable properties and applications of the GNRs.

The remainder of the paper is organized as follows. In Sec.~\ref{sec:theory}, we describe the systems under consideration, i.e., zGNRs in the presence of 5-8-5 DVs. To do so, a TB model Hamiltonian for 2D graphene lattice is presented and then the theoretical framework is introduced to calculate RKKY using the real-space Green's function. In Sec.~\ref{sec:results}, we discuss our numerical results of the exchange interaction and electronic properties of zGNRs in the presence of 5-8-5 DVs. Finally, we summarize our findings in Sec.~\ref{sec:summary}.

\section{Theory and model}\label{sec:theory}
As mentioned above, the divacancy is one of the most abundant and most important defects in  crystalline materials.
We, therefore, consider DVs produced by the removal of two neighbor carbon atoms ( the representing $p_z$ orbitals in the tight-binding model), the so-called 5-8-5 defects, so that the two sublattices are balanced~\cite{palacios_jfr_prb2008}. It should be noted that we ignore the lattice distortion.

\begin{figure*}[thpb]
      \centering
\includegraphics[width=11.2cm]{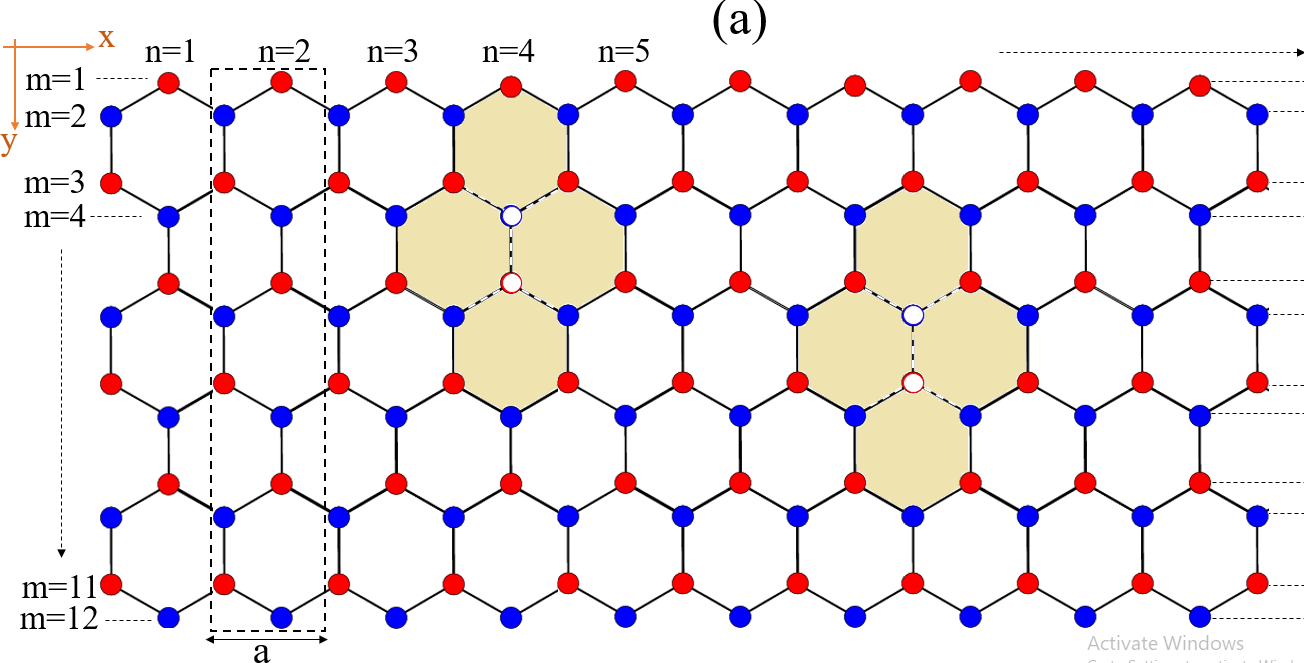}
\includegraphics[width=5.4cm]{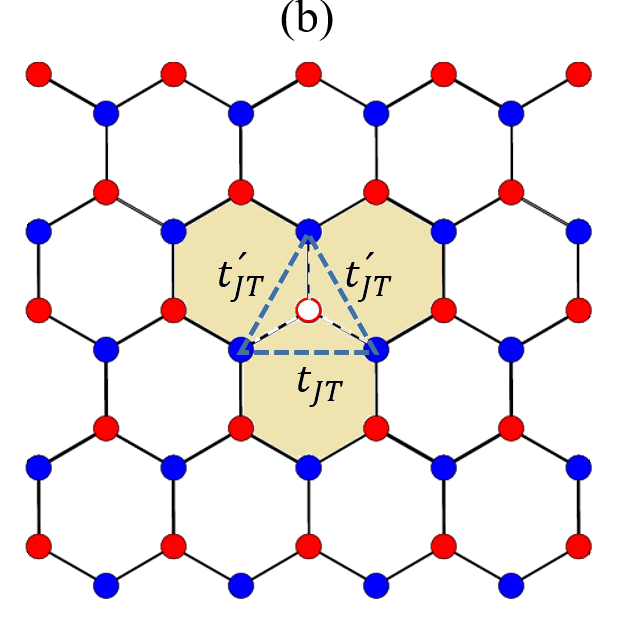}
\caption{\label{fig:schem}  (Color Online) Left panel (a): Schematic illustration of a zGNR including two 5-8-5 DVs, where the defect regions are denoted by yellow color. The dashed rectangle represents the pristine unit cell.
The length $(L=10)$ and the width$ (W=6)$ of the nanoribbon are defined as the number of the unit cells and the number of zigzag lines across the ribbon width, respectively.
Here, the location of the first and second DVs are as $(n_{1},m_{1})=(4,4)$ and $(n_{2},m_{2})=(8,6)$, respectively. For the definition of $(n_{i},m_{i})$, please see the text. Right panel (b): Schematic illustration of the hopping parameters $t_{JT}$ and $t^\prime_{JT}$ between the $sp^2\sigma$ orbitals on the carbon triangle adjacent to the vacancy, which have been created due to the Jahn-Teller distortion of the triangle.
}\label{fig:schem1}
\end{figure*}

The schematic picture of a zGNR with two DVs, where the defects regions are denoted by gray color, is shown in figure \ref{fig:schem1} (left panel). The dashed rectangle represents the pristine unit cell.
Following the conventional notation~\cite{Wakabayashi2010AdvMater,moslem-si1,MoslemBP2,moslem-si2,MoslemBP1,MoslemBS1}, the length $(L)$ and the width$ (N)$ of the nanoribbon with zigzag shaped edges on both sides are defined as the number of the unit cells and the number of zigzag lines across the ribbon width, respectively.
As shown in this figure, each atom is labeled with a pair of numbers $(m,n)$, which $m,n$ represent the $x$ and $y$ coordinates of the lattice points.

According to this notation, the positions of the two magnetic impurities are labeled by $(n_{s_1},m_{s_1})$ and $(n_{s_2},m_{s_2})$, with $(m,n)$ indices of sublattices and accordingly, the location of each divacancy defect is labeled by $(n_{i},m_{i})$ in which $n_{i}$ is the unit cell number of the $i$th DV and $m_{i}$ is the position of the upper vacancy in $i$th DV. Here, the location of the first and second DVs are $(n_{1},m_{1})=(4,4)$ and $(n_{2},m_{2})=(8,6)$, respectively.

The TB Hamiltonian for the itinerant electrons in graphene is given by \cite{cas09}:
\begin{eqnarray}
\label{H1}
{\cal H}_{NNN}=&-&t \sum_{\langle i,j \rangle,\sigma}
\left(a_{\sigma,i}^{\dag} b_{\sigma,j}
+ {\rm h.c.} \right)
\nonumber
\\
  &-& t'
\sum_{\langle \langle i,j \rangle \rangle,\sigma}
\left(a_{\sigma,i}^{\dag}a_{\sigma,j} +
   b_{\sigma,i}^{\dag}b_{\sigma,j} + {\rm h.c.} \right) \, ,
\end{eqnarray}
Here, $a_{i,\sigma}$ ($a^{\dag}_{i,\sigma}$) is the annihilation (creation) operator for a
particle with spin $\sigma$ ($\sigma = \uparrow,\downarrow$) on site ${\bf R}_i$ on sublattice A (an equivalent definition is used for $b_{i,\sigma}$ ($b^{\dag}_{i,\sigma}$) on sublattice B). The subscripts $\langle i,j \rangle$ and $\langle \langle i,j \rangle \rangle$, denote  the nearest-neighbor (NN) and next-nearest-neighbor (NNN) pairs of atoms, respectively, with $t$, the NN hopping energy between different sublattices.
The value of the NNN hopping integral $t'$ (hopping between the same sublattices) is not well-known but {\it ab initio} calculations \cite{Retal02} predicts $0.02 t \lesssim t' \lesssim 0.2t$ depending on the type of the TB parameterization. A TB fit to cyclotron resonance experiments~\cite{Deacon2007prb} suggests $t'\approx 0.1$ eV.

Based on the dedicated measurements of the DOS in graphene, by using high-quality capacitance devices, the NN and NNN terms are obtained as $t\approx 3$ eV and $t'=-0.3$ eV, respectively~\cite{Kretinin2013}.

From the density functional calculations using the linear muffin-tin orbital method and linear augmented plane wave
method, the tight-binding hopping integrals with the signs chosen such that $ t, t' > 0$ were obtained as $t \approx 2.91$ eV and $t' \approx 0.16 $ eV, respectively for graphene~\cite{brk-graphene}).

Density functional calculations plus the all-electron spin-polarized linear augmented plane wave  formalism~\cite{NandaNJP12} show that the three $sp^2 \sigma$ dangling bonds adjacent to the vacancy introduce localized states (V$\sigma$) in the mid-gap region, which split due to the crystal field and a Jahn-Teller distortion,  while the $p_z \pi$ states introduce a sharp resonance state  (V$\pi$) in the band structure.
This simple model~\cite{NandaNJP12} suggests a Jahn-Teller distortion of the carbon triangle surrounding the vacancy.
The vacancy site was modeled by simply removing a lattice site, corresponding to the vacancy potential $U_0 = \infty$. In a real material, however, $U_0$ is a large value.
Translational symmetry is broken by the presence of localized defects such as vacancies and impurities.
The vacancy has been modeled by adding an on-site perturbation $V$ to the unperturbed NN TB Hamiltonian
${\cal H}_0=-t \sum c^{\dagger}_{i\alpha}c_{j\beta} + H. c.$, with the Greek subscripts $i\alpha$ indicating the sublattice indices, so that
\begin{equation}
{\cal H}_{vcancy} =
{\cal H}_0+ V,
\label{hamil}
\end{equation}
where the localized form of the impurity potential can be written as
\begin{equation}
 V=U_{0A} c^{\dagger}_{0A}c_{0A}+U_{0A} c^{\dagger}_{0B}c_{0B},
\label{imp}
\end{equation}
 where $U_{0A}$ $(U_{0B})$ is the strength of the potential due to the vacancy on the sublattice $A$ ($B$). This
simple model suggests that the hoppings to the vacant sites are forbidden.

In graphene, the $sp^2\sigma$ states are removed away from $E_{\rm F}$ due to the strong interaction with neighboring orbitals along the C-C bonds. However, with a vacancy present, the three $sp^2\sigma$ orbitals of the three NN carbon atoms with their lobes pointed towards the vacancy have their usual bonding partners missing, so that they occur near $E_{\rm F}$, with their on-site energies $\epsilon_\sigma$ slightly below the $\pi$ orbital energies because of the $s$ orbital component present in the $\sigma$ states.


The crystal field splitting, however, can also lift the 3-fold degeneracy of the ground state into a double degenerate state and a single degenerate state. The remaining degeneracy of the double degenerate state is lifted in the presence of the Jahn-Teller distortion of the triangle, which is described by the unequal hopping $ t_{JT} \neq t^\prime_{JT}$.
Then two of the three hopping terms are modified into $t'$ as indicated in Fig. \ref{fig:schem1} (right panel).
From the DFT band structure fitting and taking into account the 2NN hopping  $t$ between the three dangling bonds in the undistorted triangle, the two unequal hopping parameters are obtained as $t_{JT} \approx 1.6 \ $ eV, and $t'_{JT} \approx 1.2 \ $ eV.

\subsection{RKKY interaction in defected zGRNs}\label{sec:RKKY}
To study the magnetic interaction between two local moments in the system, we consider the indirect exchange coupling between magnetic impurities as an RKKY form, mediated by the conduction electrons.
Using a second-order perturbation~\cite{Ruderman,Kasuya,Yosida}, the effective magnetic interaction between two magnetic moments at positions
${\bf r}_i$ and ${\bf r}_j$, is given by~\cite{Solyom,white},
\be\label{RKKY}
   J_{\textrm{RKKY}}  =  \frac{\lambda^2S(S+1)}{4 \pi S^{2}} \int d\omega
   f(\omega) \textrm{Im} \left[ G( {{\bf r}_j} ,{{\bf r}_i}, \omega)
   G({\bf r}_i,{\bf r}_j,\omega)\right]
   \ee
where $S$ is the magnitude of the impurity spin, $\lambda$ is the coupling constant between the on-site impurity spins and the spin of the itinerant electrons and $ f(\omega)= [e^{(\omega -\mu)/T} + 1]^{-1}$ is the Fermi-Dirac distribution
function at energy $\omega$, temperature $T$, and chemical potential $\mu$.
Note that we use units such that $a=1$ and $\hbar=1$ in all calculations. Making use of the Lehman representation of the Green's function,
%
the eigenfunctions $E_n$ and the wave functions $\psi_n({\bf r}_i)$ can be obtained by diagonalizing the real space Hamiltonian of zGNRs with vacancies
\be
{\cal H}_{\textrm{Deffected zGNR}}={\cal H}_{NNN}+V
\label{eq:Htot}
\ee
where ${n}$ denotes the band index and $i$ and $j$ are the carbon site index of magnetic impurities which are located at position ${\bf r}_{i}$ and ${\bf r}_{j}$.
Using the appropriate spectral functions in the low-temperature limit, the integration over energy in Eq.  (\ref{RKKY}) is therefore
\begin{equation}
    J_{\textrm{RKKY}} = -\textrm{Re}\int_{\eps<\mu} d\eps \int_{\eps'>\mu} d\eps'
    \frac{ \langle i|\delta(\eps-H)|j\rangle  \langle j|\delta(\eps'-H)|i\rangle}{\eps-\eps'},
    \label{eq:J-KPM}
\end{equation}
Finally, after straightforward calculations, the normalized RKKY interaction can be expressed in the following desired result~\cite{moslem-si1,MoslemBP2,moslem-si2,MoslemBP1,MoslemBS1}
\begin{eqnarray}
J_{\textrm{RKKY}}({\bf r}_i,{\bf r}_j) &&=-\sum_{\substack{n, n'}}[ \frac{f(E_{n})-f(E_{n'})}{E_{n}-E_{n'}}\nonumber\\
&&\times \psi_{n}({\bf r}_i)\psi^{* }_{n}({\bf r}_j)\psi_{n'}({\bf r}_j)\psi^{*}_{{ n'}}({\bf r}_i)].
\label{chiE}
\end{eqnarray}
This result, which is main equation in the present work, is a well-known formula in the linear response theory.

\section{numerical results and discussions}\label{sec:results}

In this section, we present our main results for the RKKY exchange coupling in DVs zGNRs. We evaluate the static spin susceptibility using Eq. (\ref{chiE}) in real-space for various configurations of the MIs and DVs defects.
We consider the same strength of the vacancy potential on the sublattices $A$ and $B$ i.e., $U_{0A}=(U_{0B})=U_{0}$.

\begin{figure*}[thpb]
      \centering
\includegraphics[width=18cm]{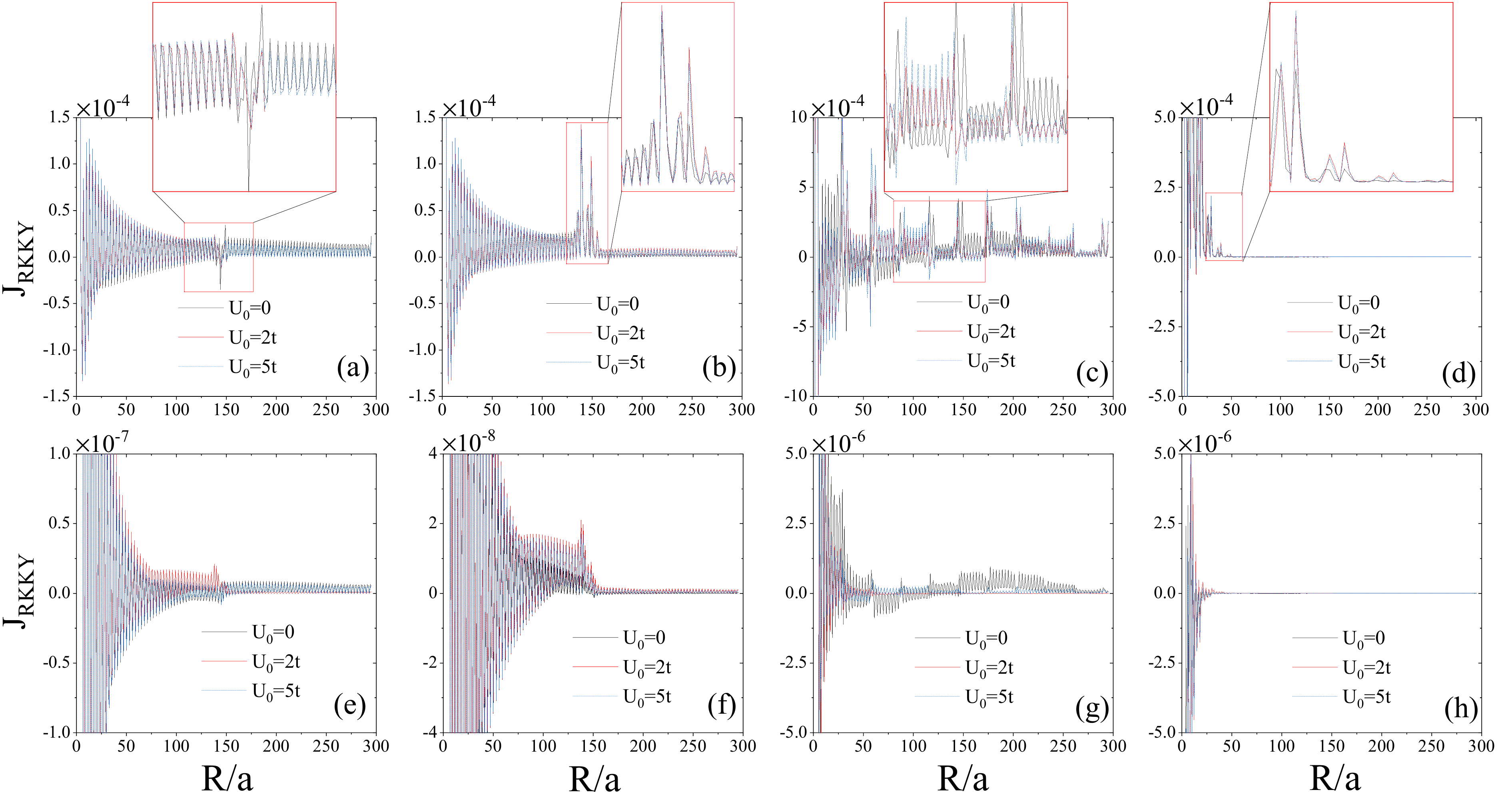}
\caption{(Color Online) Scaled RKKY interaction as a function of the impurity distance for a zGNR with $M=20,N=300$ for different strengths of the impurity potential $U_0/t=$ 0, 2, and 5.
Both the magnetic impurities are located on a same edge at positions $(5, 1)$ and $(n,1)$, with $n=6, 7, 8,...$, in the panels (a), (b) and (c),
and both are located inside the zGNR on the sublattices $(5,10)$ and $(n, 10)$, with $n=6, 7, 8,...$, in the panels (d), (e) and (f).
The panels (a), (e) are for a zGNR with one divacancy at position $(150, 4)$,
(b), (f) are for a zGNR with 2 DVs at positions $(145, 4)$ and $(155, 4)$,
(c), (g) are for a zGNR with 11 DVs at positions $(5, 4),(34, 4), (63, 4)....(295, 4)$ (DV defects, with a period of $\Delta R_{DV}/a=30$),
and finally, the panels (d), (h) are for a zGNR with 30 DVs at positions $(5, 4), (15, 4), (25, 4)....(295, 4)$ (DV defects, with a period of $\Delta R_{DV}/a=10$). The insets show a zoom on the RKKY coupling around the impurity defects.}
\label{fig:chiRimp}
\end{figure*}

First, we investigate the spatial behavior of the RKKY interaction (as a function of the dimensionless distance $R/a$ ) for two MIs in a zGNR with $M=20, N=300$, for different strengths of the impurity potential $U_0/t=$ 0, 2, and 5 eV, including single, double and multiple DVs (see Fig.~\ref{fig:chiRimp}).
In the panels (a), (b) and (c) both the MIs are located on the same edge at positions $(5, 1)$ and $(n,  1)$, with $n=6, 7, 8,...$,
and in the panels (d), (e) and (f) both are located inside the zGNR, for a configuration with the first impurity at $(5, 10)$ and the second one at $(n, 10)$, with $n=6, 7, 8,....$.

The panels (a), (e) show a zGNR with one divacancy at position $(150, 4)$,
(b), (f) represent a zGNR with 2 DVs at positions $(145, 4)$ and $(155, 4)$,
(c), (g) refer a zGNR with 11 DVs at positions $(5,4),(34, 4),(63, 4)....(295, 4)$ (DV defects, with a period of $\Delta R_{DV}/a=30$),
and finally, the panels (d), (h) illustrate a zGNR with 30 DVs at positions $(5, 4),(15, 4),(25, 4)....(295, 4)$ (DV defects, with a period of $\Delta R_{DV}/a=10$, are periodically situated in the zGNR).

The effect of the presence of the DV in this figure is apparent; the magnetic RKKY coupling shows different spatial distribution for the DV zGNR with different numbers of DVs and various spatial configurations.
Not only the magnitude but also the sign of the exchange coupling could be changed by controlling the spatial distribution of both the magnetic impurities and DVs embedded in the surface of a zGNR as well as the number of the DVs.
It is worth mentioning that, regardless of whether the MIs are at edge or not, the RKKY coupling is short-range and falls off rapidly with increasing the impurity distance, for high concentration of DVs (see panels (d),(h)), for example, here for impurity distances larger than $R/a~50$, the RKKY coupling is nearly zero.
We also consider the effect of the impurity potential $U_0$ to determine how this perturbation affect on  the spatial profile of the RKKY coupling under different numbers of local DV defects.

Most importantly, as shown in Fig.~\ref{fig:chiRimp}, one can observe that a significant perturbation with the sharp peaks appears in the spatial profile of the RKKY coupling when the second MI approaches to a divacancy, regardless of whether the MIs are at edge or interior of the ribbon.
Such perturbation of the regular RKKY oscillations around the DVs is a method of directly probing the local vacancy in a zGNR through the RKKY exchange interaction.

In addition, it has been shown that when impurities are located on the edge, the magnitude of the exchange coupling is several orders of magnitude greater than that where impurities are in the bulk.
Most importantly, zGNRs with one or two DVs have the highest difference between the edge and bulk RKKY interaction, in comparison with the samples with more DVs.
For instance, when impurities are located on the edge, the magnitude of the exchange coupling is approximately four orders of magnitude greater than that result where impurities are in the bulk, for zGNRs with one and two DVs.

\begin{figure}[thpb]
      \centering
\includegraphics[width=7cm]{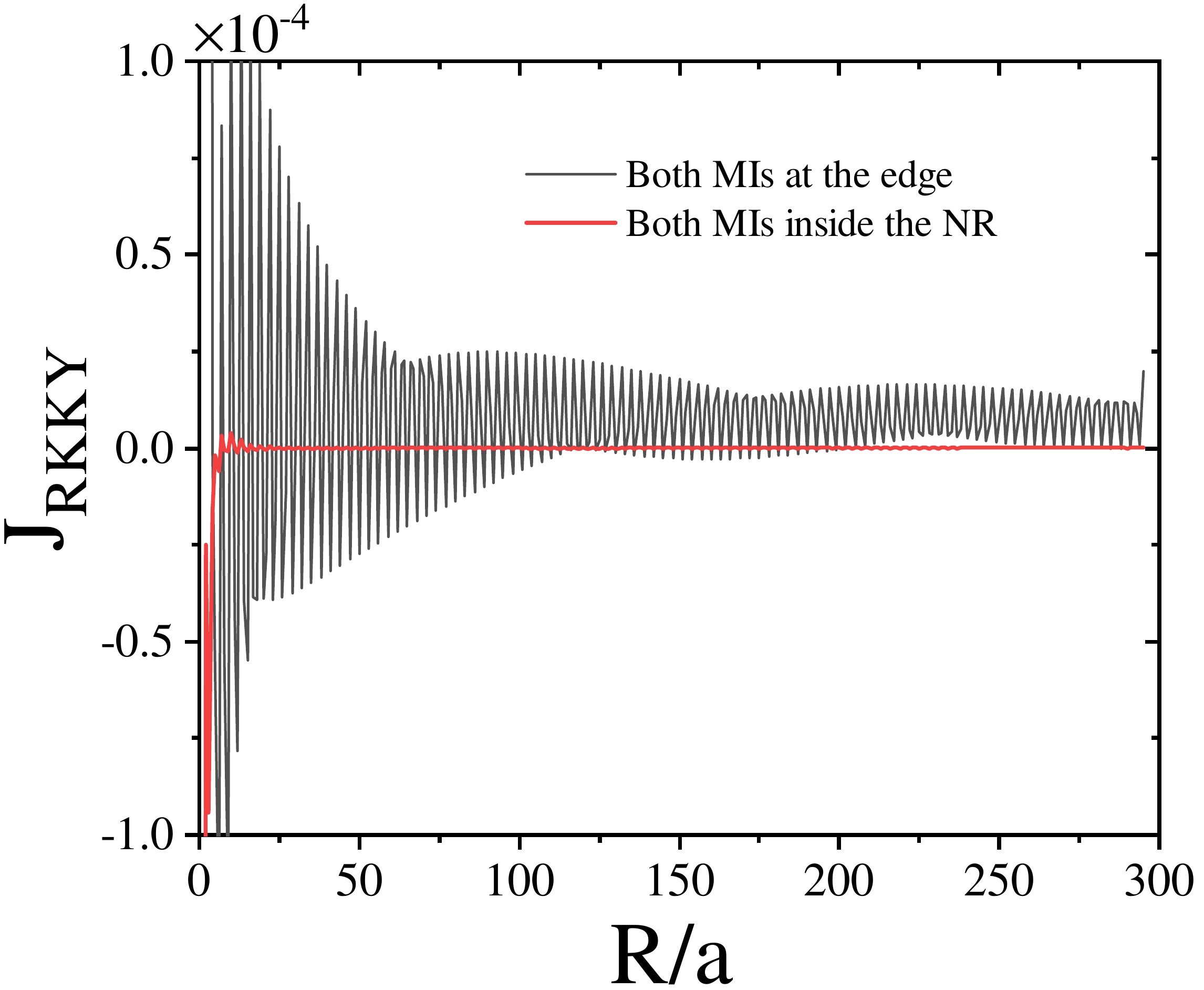}
\caption{(Color Online) Scaled RKKY interaction as a function of the impurity distance for an undefected zGNR (a pristine sample) with $M=20, N=300$, for different spatial configuration of the MIs: The red curve is for the case of both impurities located exactly at the edge at positions $(95, 1)$ and $(105, 1)$ and the black curve is for the case of both impurities located in the bulk of the zGNR at positions $(95, 10)$ and $(105, 10)$.
}\label{chi_R_0DV}
\end{figure}

To clarity, we show the result for the case of the pristine sample (an undefected zGNR) with $M=20, N=300$, for different spatial configuration of the MIs in Fig. \ref{chi_R_0DV}. The red curve is for the case when both the MIs are located on the top edge at positions $(95, 1)$ and $(105, 1)$ and the black curve is for the case when both MIs are located in the interior region of the zGNR at positions $(95, 10)$ and $(105, 10)$. The RKKY coupling shows a few oscillations in $R$, and then it decays fast with short-ranged behavior, when both the impurities are situated within the interior of the nanoribbon and there is no discontinuity in the spatial profile of the RKKY coupling.

\begin{figure}[thpb]
      \centering
\includegraphics[width=8.8cm]{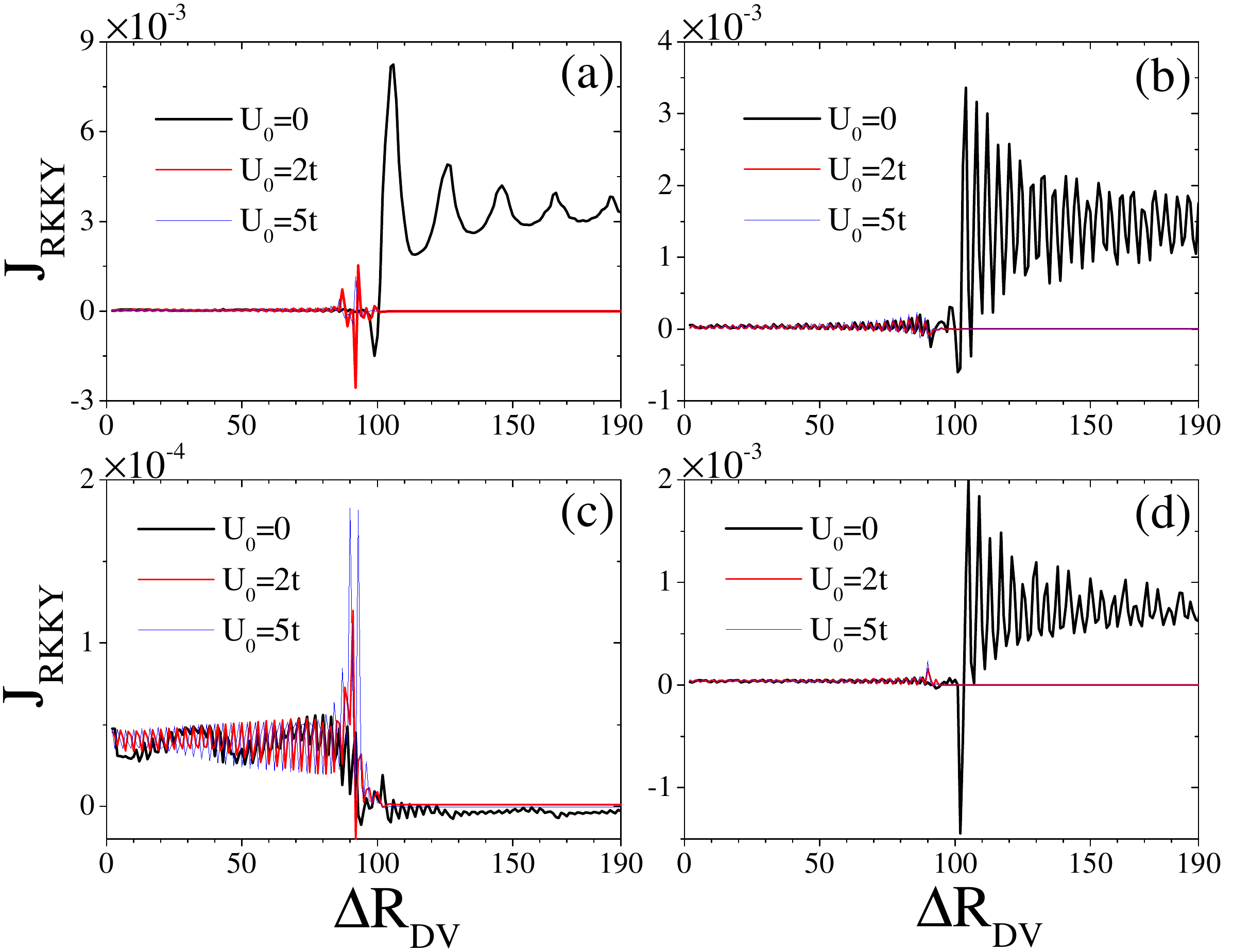}
\caption{(Color online) Scaled RKKY interaction as a function of the distance between localized DVs (in units of the unit cell length), for the intrinsic case ($E_F=0$) for different strengths of the impurity potential $U_0/t=$ 0, 2, 5 for a zGNR with $M=20, N=200$.
In the panels (a), (b) both the MIs are located on the same edge at sublattices $(95, 1)$ and $(105, 1)$ and in the panels (c), (d) both the MIs are located inside the zGNR at sublattices $(95, 10)$ and $(105, 10)$.
In the panels (a), (c) two divacancy defects are located at positions $(5, 4)$ and $(n_{2}, 4)$, with $n_{2}=6, 7, 8,....$ and in the panels (b), (d) divacancy defects are located at positions $(5, 10)$ and $(n_{2}, 10)$, with $n_{2}=6, 7, 8,....$  .}
\label{chi_R0}
\end{figure}

Here, we examine systematically how the exchange magnetic coupling of zGNR depends on the position of 5-8-5 divacancy defects (see Fig.\ref{chi_R0}). We show the range function of the RKKY interaction as a function of the distance between localized DVs (in units of the unit cell length), for a zGNR with $M=20, N=200$ (in the intrinsic case $E_{\rm F}=0$) for different strengths of the impurity potential $U_0/t=$ 0, 2, 5.
In the panels (a), (b) both the magnetic impurities are located on the same edge at positions $(95,1)$ and $(105, 1)$ and in the panels (c), (d) both are inside the zGNR on the sublattices $(95, 10)$ and $(105, 10)$.
In the panels (a), (c) two divacancy defects are located at positions $(5, 4)$ and $(n_{2},4)$ with $n_{2}=6, 7, 8,....$ (a zGNR with 5-8-5 defects placed close to the upper edge) and in the panels (b), (d) divacancy defects are located at positions $(5, 10)$ and $(n_{2}, 10)$ with $n_{2}=6, 7, 8,....$ (a zGNR with defects placed relatively farther away from the top edge).

As shown in this figure, $J_{RKKY}$ displays an oscillatory behavior with respect to the distance between two DVs.
Moreover, it is obvious that for the case when $U_0/t\neq0$, the RKKY interaction falls off very rapidly and becomes zero after $\Delta R_{DV}/a=100$. In the case when $U_0/t=0$ the trend is reverse because the RKKY coupling becomes zero for $\Delta R_{DV}/a<100$ , except in a case when DV defects placed farther away from the edge and both the magnetic impurities are located on the same edge.

\begin{figure*}[thpb]
      \centering
\includegraphics[width=16cm]{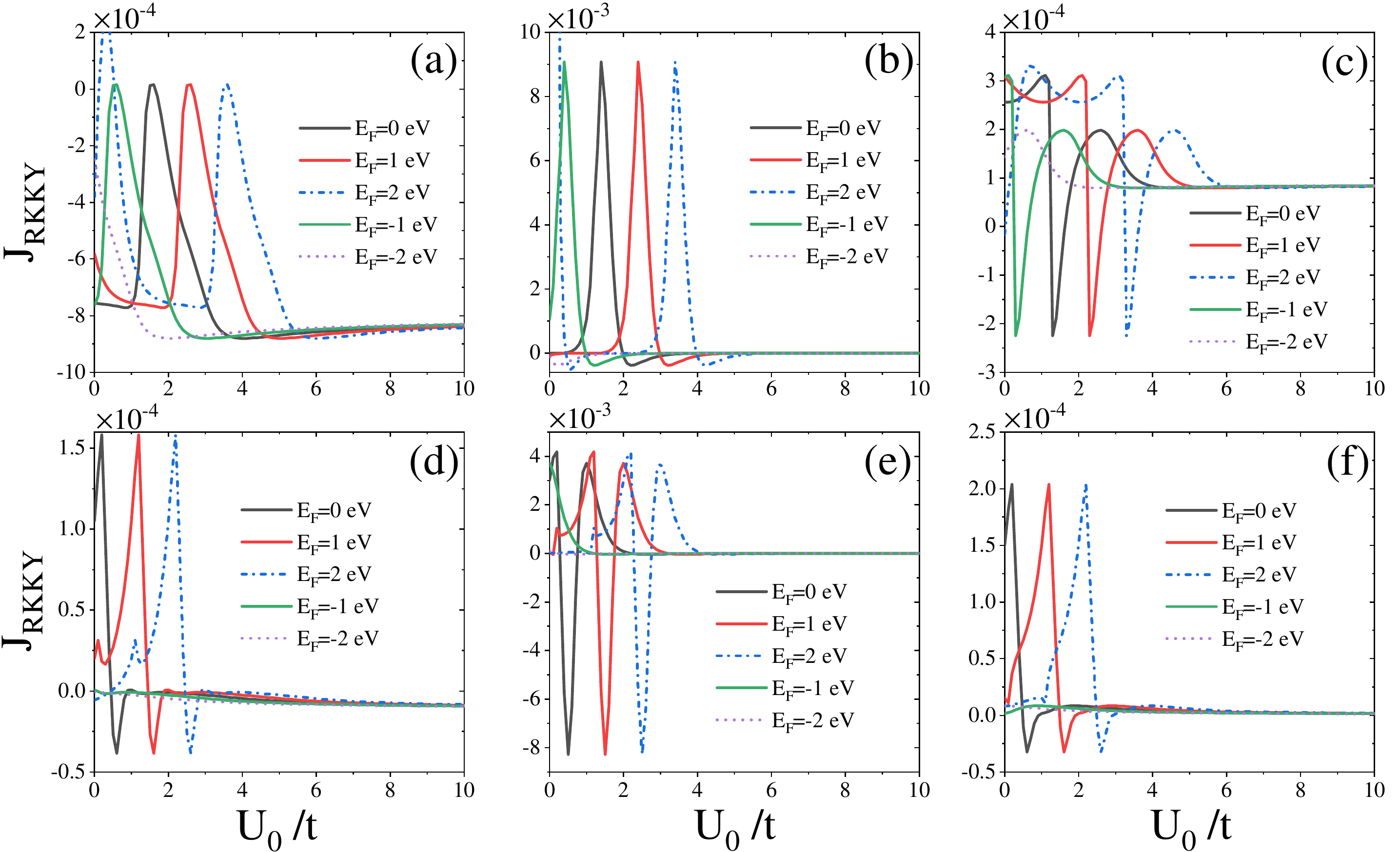}
\caption{\label{fig:schem}  (Color Online) Scaled RKKY interaction as a function of the vacancy potential $U_0/t$ for
zGNRs with $M=20,N=200$ for various Fermi energies $E_F=0,\pm 1, \pm 2$ eV. In the panels (a),(b) and (c), two DVs are located at positions $(95,4)$ and $(105,4)$ (a zGNR with 5-8-5 defects placed close to the upper edge) and in the panels (d),(e) and (f) two DVs are located at positions $(95,10)$ and  $(105,10)$ (a zGNR with 5-8-5 defects placed relatively farther away from the top edge).
In the panels (a),(d) both the MIs are located on the same zigzag edge at the sites $(95,1)$ and $(105,1)$, and in the panels (b),(e) both are located inside the zGNR on the sublattices $(95,10)$ and $(105,10)$, and finally, in the panels (c),(f) both the MIs are fixed at the counterpart zigzag edges at lattice sites $(100,1)$ and $(100,20)$.
}\label{fig:chi_U0}
\end{figure*}

We display in Fig.~\ref{fig:chi_U0} the calculated results for the scaled RKKY interaction as a function of the vacancy potential $U_0/t$, for a
zGNR with $M=20,N=200$ for various Fermi energies $E_{\rm F}=0,\pm 1, \pm 2$.
The panels (a), (b), and (c), correspond to situations when both the DV defects are placed close to the upper edge. We fix one of the DVs at the position $(95, 4)$ and the location of the second divacancy is at $(105, 4)$ and in the panels (d), (e) and (f), we consider the case when both the divacancy defects are away from the edge at positions $(95, 10)$ and $(105, 10)$.
In the panels (a), (d) both the MIs are located on the same zigzag edge at the sites $(95, 1)$ and $(105, 1)$, in the panels (b), (e) both are located inside the zGNR on the sublattices $(95, 10)$ and $(105, 10)$, and finally, in the panels (c),(f) both the MIs are fixed at the counterpart zigzag edges at lattice sites $(100, 1)$ and $(100, 20)$.

As shown, the presence of DVs profoundly alters the magnetic ground state of defected zGNR.
The quenching of the RKKY interaction at and above a certain vacancy potential is clear in these figures. Therefore, the Fermi energy and the spatial configuration of both the MIs and the DVs have a very significant impact on the vacancy potential engineering of the magnetic coupling in 2D zGNRs.
It is worth mentioning that, regardless of whether the DVs are close to the edge or not, the RKKY interaction has a peak structure with the maximum or minimum value. In all cases, the main peak shifts toward the higher vacancy potentials with an increase in the Fermi energy, for positive Fermi energies.
By contrast, in the absences of the NNN hopping $t'$, the LDOS is a symmetric function with respect to a change of sign of
the energy $E \rightarrow -E$, with a sharp peak exactly at $E=0$ (not shown here).
When the DV defect is moved toward the center of the ribbon in a double divacancy defect, makes the local DOS decreases rapidly.

\begin{figure*}[thpb]
      \centering
\includegraphics[width=16cm]{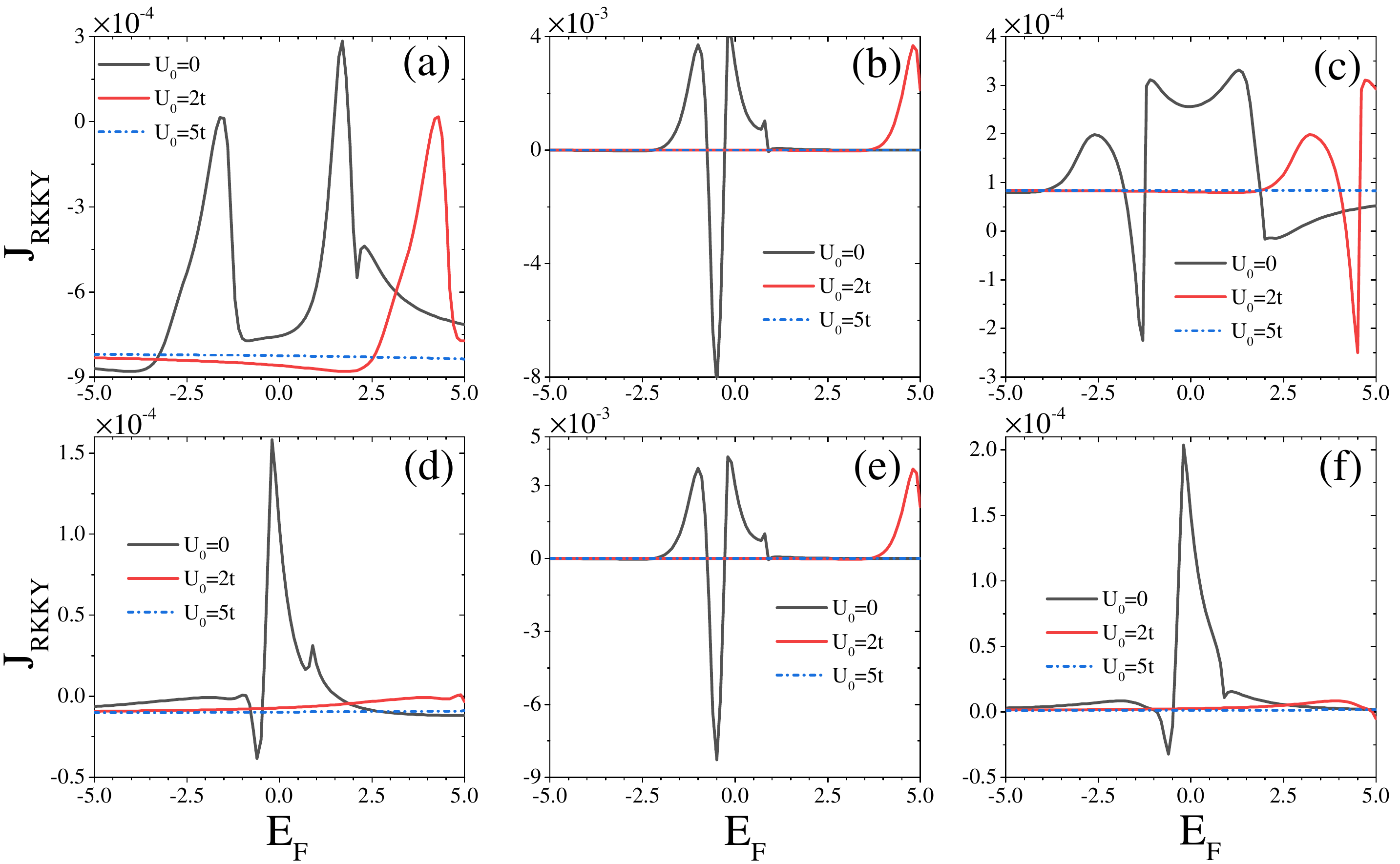}
\caption{\label{fig:schem} (Color Online) Scaled RKKY interaction as a function of the Fermi energy for zGNRs with $M=20,N=200$ for various vacancy potentials $U_0/t=0, 2, $.
The position of both the DVs and the MIs are the same as Fig.\ref{fig:chi_U0}: in the panels (a),(b) and (c), two DVs are located at positions $(95,4)$ and $(105,4)$ (a zGNR with DV defects placed close to the upper edge) and in the panels (d),(e) and (f) DVs are located at positions $(95,10)$ and  $(105,10)$. In the panels (a),(d) both the MIs are located on the same zigzag edge at the sites $(95,1)$ and $(105,1)$, and in the panels (b),(e) both are located inside the zGNR on the sublattices $(95,10)$ and $(105,10)$, and finally, in the panels (c),(f) both the MIs are fixed at the counterpart zigzag edges at lattice sites $(100,1)$ and $(100,20)$.
}\label{fig:chi_EF}
\end{figure*}

In order to further investigate of the presence of DV in the zGNR we also calculate the RKKY coupling as a function of the Fermi energy for different configurations of the MIs in Fig.~\ref{fig:chi_EF}.
The position of both the DVs and the MIs are the same as Fig.\ref{fig:chi_U0}: in the panels (a), (b) and (c), two DVs are located at positions $(95, 4)$ and $(105, 4)$, and in the panels (d), (e) and (f) DVs are located at positions $(95, 10)$ and  $(105, 10)$.
In the panels (a), (d) both the MIs are located on the same zigzag edge at the sites $(95, 1)$ and $(105, 1)$, in the panels (b), (e) both are located inside the zGNR on the sublattices $(95, 10)$ and $(105, 10)$, and finally, in the panels (c), (f) both the MIs are fixed at the counterpart zigzag edges at lattice sites $(100, 1)$ and $(100, 20)$.

We find here that the RKKY coupling depends on the DV positions and more strongly on the distance from the edge.
Now, we show that different types of vacancies in the same zGNR gives rise to different changes in the electronic and magnetic properties.
Interestingly, in the limit of high vacancy potential, the strength of the RKKY interaction is approximately unchanged in terms of the Fermi energy {\it i.e.,} the magnetic ground state of the system is constant either positive (ferromagnetic) or negative (antiferromagnetic) with varying the Fermi energy, depending on the divacancy distance from the edge as well as the position of two MIs.
In addition, in the limit of high vacancy potential, when both impurities are located inside the zGNR as shown in panels (b), (e), the RKKY coupling nearly becomes zero.

\begin{figure}[]
\begin{center}
\includegraphics[width=8.5cm]{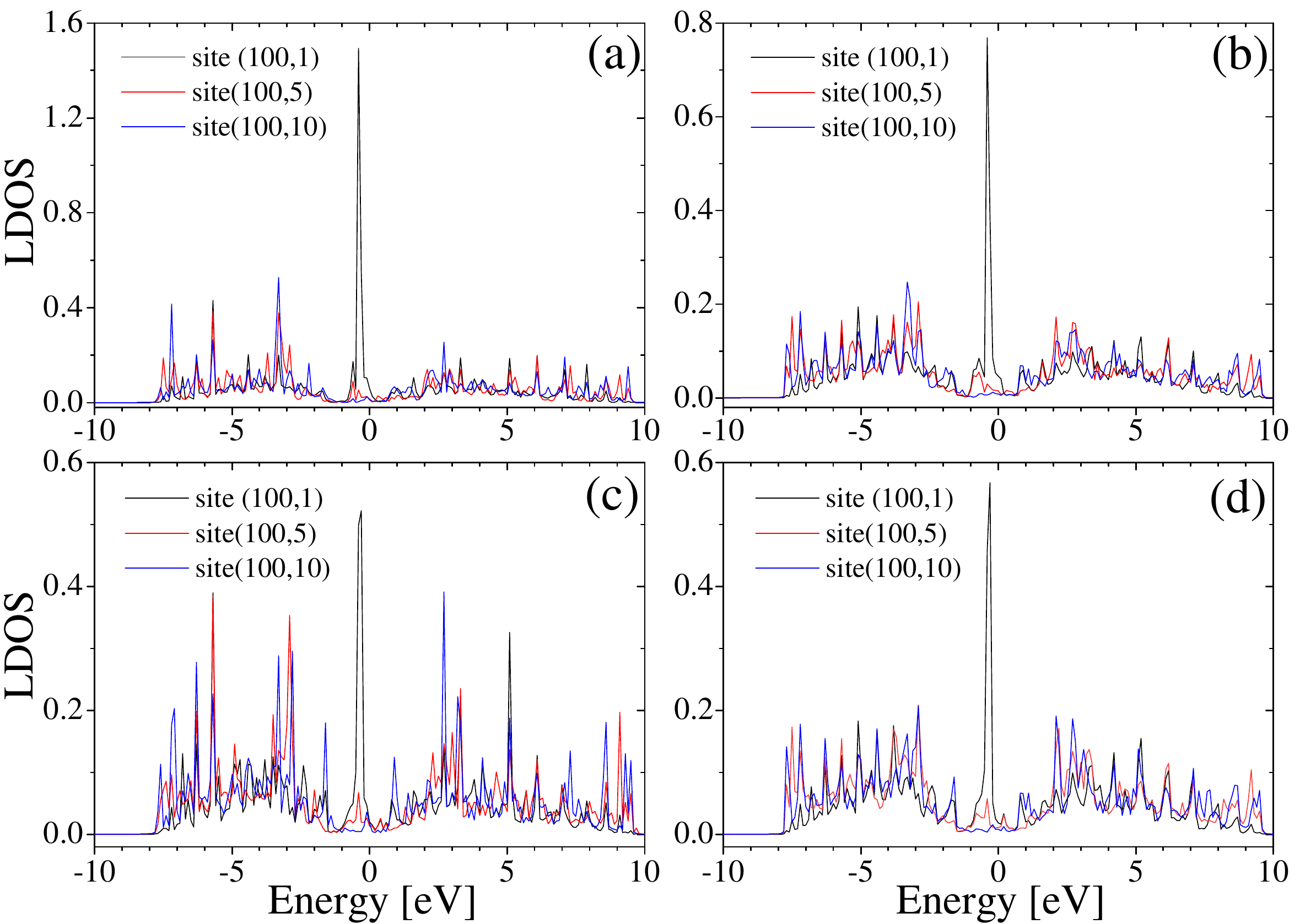}
\end{center}
\caption{\label{fig:LDOS1} (Color online) Site-resolved local densities of states (LDOS) for different lattice sites (for both the edge and bulk sublattices) for a zGNR with $M=20,N=200$ with $U_0/t=2$. The panels (a),(c) are for a zGNR with two DVs located at $(95,4)$, $(105,4)$ sites and $(95,10)$, $(105,10)$ sites, respectively and the panels (b),(d) are for a zGNR with one DV located at $(95,4)$ sites in the panel (c), and located at $(95,10)$ sites in the panel (d).
}
\end{figure}

Furthermore, to understand the effects of the position of the MIs and the DVs on the RKKY properties of zGNRs, we study the local density of state (LDOS) of the zGNRs. Corresponding site-resolved LDOS for the $i$th site, at a given position ${\bf r}$ and energy $E$, is obtained from the imaginary part of the unperturbed Green's function as
\begin{equation}\label{2}
{\text {LDOS}}({\bf r},E)=-\frac{1}{\pi} \textrm{Im} G^0({\bf r},{\bf r}; E)
\end{equation}
where the unperturbed Green’s function matrix $G^0({\bf r},{\bf r}; E)$ is expressed as
\begin{equation}\label{2}
\noindent
G^0({\bf r},{\bf r}; E)=\frac{1}{(E+i\eta)\mathbb{1}-H}.
\end{equation}
where $\eta$ is a positive infinitesimal number which is taken as 1 meV in our calculations without specification.
Now, the effects of multiple DVs located near the edge as a function of the electron energy will be discussed.
The calculated LDOS for a zGNR with $M=24, N=300$ and $U_0/t=2$ is shown in Fig.~\ref{fig:LDOS1}.
The panels (a), (c) are for a zGNR with two DVs located at $(95, 4)$, $(105, 4)$ sites and $(95, 10)$, $(105, 10)$ sites, respectively and the panels (b), (d) are for a zGNR with one DV located at $(95, 4)$ sites in the panel (c), and located at $(95, 10)$ sites in the panel (d).

The effect of the presence of the divacancy is clear;
similar to the RKKY coupling, the LDOS depends on the divacancy positions and more strongly on the distance from the edge.
The zGNRs with different number of DVs and different DV positions have different electronic LDOSs at a certain site.
When the divacancy defect is placed close to one of the edges, the LDOS becomes more sensitive to the site positions.
It is worth mentioning that, regardless of whether the DVs are close to the edge or not, the LDOS has a sharp peak for the edge sites.
In all cases, this main peak in the vicinity of the zero-energy is asymmetric because in the TB model Eq.~(\ref{H1}) we take into account the NNN $t'$ that leads to the electron-hole asymmetry and shift of the zero LDOS peak toward the lower energies.
By contrast, in the absences of the NNN hopping $t'$, the LDOS is a symmetric function with respect to a change of sign of
the energy $E \rightarrow -E$, with a sharp peak exactly at $E=0$ (not shown here).

\begin{figure}[]
\begin{center}
\includegraphics[width=8cm]{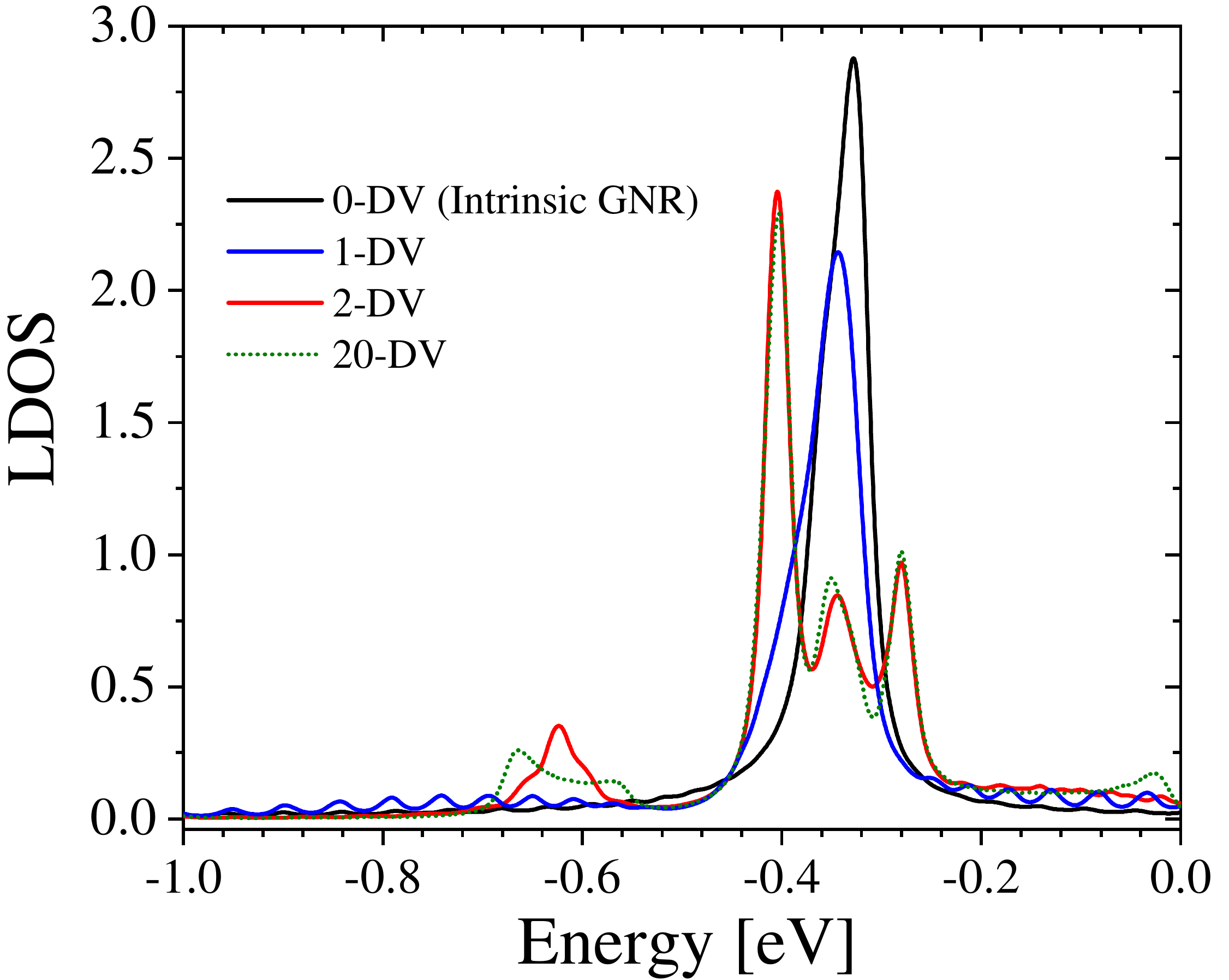}
\end{center}
\caption{\label{fig:LDOS2} (Color online) Site-resolved local densities of states (LDOS) calculated for an edge site with coordinate $(100,1)$, in a zGNR with $M=20,N=200$. The DVs with the vacancy potential $U_0/t=5$, have been located close to the upper edge ($m_{i}=4$). The corresponding LDOS for a pristine monolayer zGNR (defect-free zGNR) is shown with black-solid line.
}
\end{figure}

Numerical calculated LDOS for an edge site with coordinate $(100,1)$ in a zGNR with $M=20, N=200$, is shown in Fig. \ref{fig:LDOS2}, by considering the DVs with the vacancy potential $U_0/t=5$ located close to the upper edge ($m_{i}=4$). The corresponding LDOS for a pristine monolayer zGNR (defect-free zGNR) is shown with a black-solid line.

When defects are present within the zGNR, the resulting electronic LDOS becomes different from that for defect-free zGNR.
Comparing to the LDOS of the defect-free ribbon with a sharp peak around the $E=-0.33$ eV, but now for a zGNR with two DVs, besides the main peak at $E=-0.4$ eV three additional peaks appear; one peak at $E=-0.28$ eV and two peaks at $E=-0.34$ eV and $E=-0.62$ eV.

\section{summary}\label{sec:summary}
In summary, based on the Green’s function approach in combination with the tight-binding approximation, we have investigated the effect of 5-8-5 DVs on the electronic and magnetic properties of zGNRs. We have discussed the evolution of the RKKY interaction mediated by a background of the conduction electrons of defected zGNRs including single, double, and multiple DVs.
DVs are modeled by removing two adjacent carbon atoms from their sites where the hoppings to the vacant sites are forbidden.
The calculations show that the changes in the electronic LDOS and the exchange coupling of zGNRs depend on the location of the DVs with respect to the ribbon edges and on the number of the DVs. We have found that introducing vacancies into zGNR changes the spatial variation of the RKKY interaction, particularly for those magnetic moments located around the vacancies.
The zGNRs with one or two DVs have the highest difference between the edge and bulk RKKY interaction, in comparison with the samples with more DVs. We have shown that different values of the vacancy potential in the same zigzag nanoribbon give rise to different changes in electronic and magnetic properties.
A periodic divacancy created in a zGNR causes a dramatic change in the magnetic ground state of the ribbon. A strong perturbation of the regular RKKY oscillations appears in the spatial profile of the RKKY coupling when the magnetic impurities approach a divacancy.

Our results suggest that the defect engineering of atomic vacancies is a promising way to modify the magnetic properties of graphene nanoribbons that can lead to remarkable properties and applications in spintronics based on monolayer zGNRs.

\begin{acknowledgments}
This work is supported by the Iran Science Elites Federation.
\end{acknowledgments}

\end{document}